\documentclass[letterpaper]{JHEP}

\usepackage{amssymb,amsmath}
\usepackage{epsfig}

\def\be{\begin{eqnarray}}
\def\ee{\end{eqnarray}}

\def\3{\ss}

\def\half {\frac{1}{2}}
\def\quart {\frac{1}{4}}

\preprint{{\tt hep-th/0410046}}

\title{On the CSFT approach to localized closed string tachyons}

\author{Oren Bergman and Shlomo S.~Razamat\\
Department of Physics\\
Technion, Israel Institute of Technology\\
Haifa 32000, Israel\\
\email{bergman, razamat@physics.technion.ac.il}}

\abstract{We compute the potential for localized closed string tachyons 
in bosonic string theory on the orbifold ${\mathbb C}/{\mathbb Z}_4$ using level-truncated
closed string field theory.
The critical points of the potential exhibit features which agree with their
conjectured identification as lower-order orbifolds.
However this case also raises some questions regarding the quantitative
predictions associated with these conjectures.}


\begin{document}


\section{Introduction}

Though we've learned much about the role of tachyons in open string theory,
closed string tachyons have remained somewhat of a mystery. 
The main difficulty is that closed string tachyons
couple directly to gravity and the dilaton, so the back-reaction of
the tachyon condensate on the background is important.
In an interesting development, Adams, Polchinski and Silverstein (APS) proposed
a scenario for the condensation of {\em localized} closed string tachyons
which appear in the twisted sectors of the orbifold $\mathbb C/\mathbb Z_N$ in Type II
superstring theory \cite{APS}.
This describes a cone with deficit angle $\theta_N=2\pi(1-1/N)$,
and the tachyons in question live at the tip of the cone.
They conjectured that the condensation of the twisted tachyons 
affects the background only locally by flattening the cone.
The minimum should correspond to flat space, and the other critical
points to cones with a smaller deficit angle, namely $\mathbb C/\mathbb Z_M$
with $M<N$. 
The conjecture was supported both by a D-brane probe analysis
\cite{APS,Michishita} and by worldsheet renormalization group techniques \cite{APS,Vafa,HKMM,HMT}.
This background has also been studied from the low-energy supergravity
point of view, and was shown to produce radial gravity and dilaton waves 
which can gradually reduce the deficit angle \cite{GH,Headrick}.
A large $N$ approach to the problem has been discussed in \cite{Sarkar,Dabholkar2}.
The relation of localized closed string tachyons to semiclassical
instabilities has been addressed in \cite{flournoy}.

In a recent paper Okawa and Zwiebach have set out to test the APS conjecture
in the context of the bosonic string
using closed string field theory \cite{OZ}.
In this case a bulk (untwisted) tachyon exists for all $N$, but one considers
the condensation of just the localized tachyons, which include the twisted
tachyons as well as localized modes of the untwisted tachyon.
This is similar in spirit
to the discussion of open string tachyon condensation in theories containing
also closed string tachyons.

Computation of the tachyon effective potential in open string field
theory using level-truncation has lead to several remarkable tests of 
Sen's conjectures \cite{Sen_conjectures,Sen_review} regarding open string 
tachyon condensation (for a review
see \cite{Taylor}).
In trying to extend these techniques to closed string tachyons in closed
string field theory one encounters two immediate problems.
The first is the lack of a clear quantitative prediction.
In the open string case Sen's conjecture can be translated into 
a simple numerical prediction for the tachyon potential,
namely that the depth of its minimum is equal to the tension of
the original unstable D-brane.
Dabholkar has proposed an analogous prediction for the twisted tachyons of 
$\mathbb C/\mathbb Z_N$
\cite{Dabholkar}. From the effective action on the orbifold,
\be\label{grav_action}
S=-\frac{1}{2\kappa_{N}^2}
\int_{{\mathbb R}^{1,7}\times {\mathbb C}/{\mathbb Z}_N} d^{10}x\sqrt{-g}e^{-2\Phi}R-
\int_{{\mathbb R}^{1,7}}d^{8}x\sqrt{-g^{(D-2)}}e^{-2\Phi}V_N(T)\;,
\ee
and assuming that the potential for the twisted tachyons
$V_N(T)$ is independent of the metric, one gets a relation between the
potential and the deficit angle 
\be
 \theta = \kappa^2_{N} V_N(T) \;.
\label{prediction_1}
\ee
This means that the relative depths of the various critical points of
the potential are given by the difference in deficit angles of the 
orbifolds they represent. Expressed in terms of the normalized 
dimensionless potential
\be\label{f_def}
 \widetilde{f}_N(T) \equiv {\kappa^2_{N}V_N(T)\over 2\pi\left(1-\frac{1}{N}\right)}\;,
\ee
the prediction for the critical point corresponding to $\mathbb C/\mathbb Z_M$
is therefore
\be 
 \widetilde{f}_N(T_M) = -{{1\over M}-{1\over N}\over 1-{1\over N}} \qquad M=1,\ldots, N-1 \;.
\label{prediction_2}
\ee
This is also consistent with the fact that the ADM mass of a 
co-dimension 2 object, like the $\mathbb C/\mathbb Z_N$ orbifold 7-plane, 
is given by its deficit angle.
However the actual potential depends also on localized modes of the
untwisted (bulk) tachyon, and therefore it depends on the metric 
through the metric dependence of the bulk tachyon expectation values
\cite{OZ}. Strictly speaking therefore, the prediction (\ref{prediction_1}) only
holds to lowest level, where the massless (level 2) fields do not appear.
It does however need to be refined by including, and then integrating out, 
localized modes of the
bulk tachyon. It is not clear how to generalize the prediction to include
metric dependence and higher-level fields.  
The second problem is that, unlike open string field theory, 
closed string field theory has an infinite number 
of interaction vertices, so that one would have to include an infinite
number of terms in the action at every level \cite{Zwiebach_CSFT}. 
In fact, very little is known about the strength of the interactions beyond
the quartic vertex (for which we only have a numerical approximation 
\cite{Belopolsky,Moller}).

Given these problems, it is quite remarkable how close one gets 
using a lowest-level truncation and only the cubic vertex \cite{OZ}.
Okawa and Zwiebach computed the localized tachyon potential 
for $\mathbb C/\mathbb Z_2$ and $\mathbb C/\mathbb Z_3$ and found an 
agreement of about 35\%
with the predicted depth of the minimum of the potential in both 
cases.\footnote{An agreement of about 70\% was reported in 
\cite{OZ} due to an error in identifying the orbifold gravitational
coupling $\kappa^2_N$ with the flat space gravitational coupling $\kappa^2$.
This error was also present in an earlier version of this paper.} 
For $\mathbb C/\mathbb Z_3$ there is a second critical point, conjectured to 
correspond
to the decay to $\mathbb C/\mathbb Z_2$. Its depth was found to be 44\% of the 
predicted
value. It was also shown that including the quartic vertex does not change 
the qualitative features of the potential, although it changes the depths 
of the critical points somewhat. This suggests that perhaps the higher
vertices should be assigned an intrinsic level, and should therefore 
be truncated according to the level of the approximation.

In this paper we present further tests of this approach to localized 
tachyon condensation by analyzing the lowest-level localized tachyon potential
of $\mathbb C/\mathbb Z_4$. 
In this case there are three predictions corresponding
to the three possible decay products: flat space, $\mathbb C/\mathbb Z_2$ and
$\mathbb C/\mathbb Z_3$. 
The critical points of the potential are arranged
on an (irregular) tetrahedron in the three-dimensional space of twisted tachyons
$(t, t')$, where $t$ is complex and $t'$ is real.
The maximum, $\mathbb C/\mathbb Z_4$, is at the body-center, and the four 
degenerate minima are at the vertices. The minimum
is found at 25\% of the predicted value for flat space.
The $\mathbb C/\mathbb Z_3$ points are located at the four faces
of the tetrahedron, and give 46\% of the predicted value.
The $\mathbb C/\mathbb Z_2$ points are located at the 
six edges, and are actually split into a group of four and a group
of two. The former give 39\% of the predicted value, and the latter
34\%. 
Technically, the split is due to the fact that the potential does
not respect the full tetrahedral symmetry, but only a $\mathbb Z_4$
(or really $D_4$) subgroup, which is the quantum symmetry of the orbifold.
It is not clear to us whether the tetrahedral symmetry should be restored at higher level.

In section 2 we review some of the necessary ingredients needed for computing
orbifold tachyon potentials. In section 3 we
compute the localized tachyon potential for the bosonic string on 
$\mathbb C/\mathbb Z_4$, and in section 4 we find the critical points 
and compare their depths with the predicted values.
Section 5 contains our conclusions.
We have also included two appendices which contain some details
of the computation.


\section{Necessary ingredients and general strategy}

Let us first review briefly the necessary ingredients for computing
localized tachyon potentials in ${\mathbb C}/{\mathbb Z}_N$ in general.
This orbifold has $N-1$ twisted sectors corresponding to $N-1$
twist fields $\sigma_k$ in the orbifold CFT. The conformal dimensions
of $\sigma_k$ are $h_k=\bar{h}_k = {1\over 2}{k\over N}\left(1-{k\over N}\right)$,
so the fields $\sigma_k$ and $\sigma_{N-k}$ are paired up.
The orbifold CFT has a quantum symmetry $\mathbb Z_N$ under which the twist fields
transform as 
\be 
 \sigma_k \rightarrow \exp\left({2\pi ki\over N}\right) \sigma_k \;,
\ee
and all correlation functions are invariant under this.
The untwisted operators ${\cal U}_p$ carry momentum $p$ along the 
directions of the cone 
and must be projected to the invariant sector:
\be
 {\cal V}_p = {1\over N} \sum_{k=0}^{N-1}{\cal U}_{\alpha^k p} \;,
\ee 
where $\alpha = \exp(2\pi i/N)$. 
The basic correlation functions are given by\footnote{We set $\alpha'=1$.}
\cite{DFMS}
\begin{eqnarray}
 \langle\sigma_{N-k}(\infty)\sigma_k(0)\rangle &=& V_{D-2} \nonumber\\
\label{correlators} 
\langle{\cal V}_p(\infty)\sigma_{N-k}(1)\sigma_k(0)\rangle &=& 
\delta^{- p^2/4}\, V_{D-2} \\
 \langle\sigma_{N-2k}(\infty)\sigma_k(1)\sigma_k(0)\rangle &=&
\sqrt{{\tan(\pi k/N)\over 2\pi^2}}\,
{\Gamma^2\left(1-{k\over N}\right)\over \Gamma\left(1-{2k\over N}\right)}\nonumber\,
V_{D-2} \;,
\end{eqnarray}
where $\delta$ is the following function of $k/N$:
\be
 \delta\left({k\over N}\right) = 
   \exp\left[2\psi(1) - \psi\left({k\over N}\right) 
     - \psi\left(1-{k\over N}\right)\right] \;,
\ee
and $\psi(x) = \Gamma'(x)/\Gamma(x)$. Some useful values of this function are
$\delta(1/2)=2^4$, $\delta(1/3)=3^3$ and $\delta(1/4)=2^6$.
The generalization to include momentum in the seven directions transverse
to the cone is straightforward, but does not interest us here
since we focus on uniform (in the seven transverse dimensions) 
tachyon condensation.

The string field theory action to cubic order is given by
\be 
 S = - 2 \sum_{\alpha,\beta}
      (h_\beta-1)\phi^\alpha m_{\alpha\beta} \phi^\beta
     - {2\kappa\over 3!} \sum_{\alpha,\beta,\gamma}
     {\cal R}^{6-2(h_\alpha+h_\beta+h_\gamma)}
     \phi^\alpha \phi^\beta \phi^\gamma C_{\alpha\beta\gamma}\;,
\label{SFT}
\ee
where $\kappa$ is the (square root of the) flat space gravitational coupling constant,
$\phi^\alpha$ are the component fields of the string field
$|\Psi\rangle = \sum_\alpha c_1 \overline{c}_1 \phi^\alpha |{\cal O}_\alpha\rangle$,
and $m_{\alpha\beta}$ and $C_{\alpha\beta\gamma}$ are given by 
\begin{eqnarray}
 m_{\alpha\beta} &=& \langle bpz({\cal O}_\alpha)|
 c_{-1}\overline{c}_{-1}c_0^- c_0^+ c_1\overline{c}_1|{\cal O}_\beta\rangle \\
 C_{\alpha\beta\gamma} &=& \langle c\overline{c}{\cal O}_\alpha(0)\,
    c\overline{c}{\cal O}_\beta(1)\,
    c\overline{c}{\cal O}_\gamma(\infty)\rangle \;.
\end{eqnarray}
The number ${\cal R}$ is the inverse of the mapping radius for the map from the
punctured unit disk to a punctured $120^\circ$ wedge of the complex plane,
and is given by
\be
 {\cal R} = {3\sqrt{3}\over 4} \;.
\ee
In particular the action for the bulk tachyon with momentum
in the cone directions $p$, $u(p)c_1\overline{c}_1|p\rangle$, is given by
\begin{eqnarray}
 S_u &=& -{1\over 2} \int {d^2p\over (2\pi)^2} u(-p)(p^2 - 4) u(p) \nonumber\\
 && - {4\kappa\over 3!}\int \prod_{i=1}^3 \left[
   {d^2 p_i\over (2\pi)^2} {\cal R}^{2-{1\over 2}p_i^2} u(p_i) \right]
   (2\pi)^2 \delta^{(2)}(p_1 + p_2 + p_3) \;.
\label{bulk_action}
\end{eqnarray}
This result is universal for all $N$. The integrals 
are over all momenta in $\mathbb{C}$, where one identifies
$u(p)=u(\alpha^k p)$.
The rest of the tachyon action involves the twisted tachyons and is computed
using the correlation functions in (\ref{correlators}). 
We will work in the lowest-level approximation, in which we keep only 
the ground state tachyons in each twisted sector, and include bulk
tachyon modes up to a level equal to the level of the highest twisted
tachyon. Since the level of the bulk tachyon is given by\footnote{The level
is defined as $\ell\equiv L_0 + \overline{L}_0 + 2$.} $\ell_u(p) = p^2/2$,
this means that
the momentum integrals will be truncated.

The gravitational coupling constant in flat space $\kappa^2$ is related
to the orbifold gravtitational constant $\kappa^2_N$ as
\be\label{Newton_rel}
\kappa_{N}^2=\frac{\kappa^2}{N}.
\ee
One way to see this is to note that the integration region in the
orbifold effective theory (\ref{grav_action}) is the the reduced space 
$\mathbb{C}/{\mathbb Z}_N$, whereas in the string field theory action 
(\ref{bulk_action}) it is the covering space $\mathbb{C}$.
The relative factor of $N$ in (\ref{Newton_rel}) comes from the 
relative volumes of the two spaces.\footnote{We would like to thank the referee 
of this paper for turning our attention to this point, and to Yuji 
Okawa and Barton Zwiebach for this simple explanation. The relation 
(\ref{Newton_rel}) is also implied in \cite{Headrick_Rae}}. 






\section{The Tachyon potential in ${\mathbb C}/{\mathbb Z}_4$}

The conformal dimensions of the three twist fields are given by
\be
h_{\sigma_1}=h_{\sigma_3}=\frac{3}{32} \qquad h_{\sigma_2}=\frac{1}{8}\;.
\ee
This gives a complex tachyon $t$ of mass squared $m^2_t = -29/8$
and level $\ell_t = 3/16$, and a real tachyon $t'$ of mass squared
$m^2_{t'} = -7/2$ and level $\ell_{t'} = 1/4$.
The lowest-level localized tachyon potential to cubic order is then given by
\be
 V_4(u,t,t') = V_{u^2} + V_{u^3} + V_{t^*t} + V_{t^{\prime 2}}
   + V_{ut^*t} + V_{ut^{\prime 2}} + V_{(t^2 + t^{*2})t'} \;,
\ee
where $V_{u^2}$ and $V_{u^3}$ are given by (\ref{bulk_action}), and the rest
are given as follows
\begin{eqnarray}
 V_{t^*t} &=& -{29\over 8} t^* t \\
 V_{t^{\prime 2}} &=& -{1\over 2}{7\over 2} t^{\prime 2} \\
 V_{ut^*t} &=& 4\kappa{\cal R}^{45/8} t^* t \int{d^2p\over (2\pi)^2}
    \left[{\cal R}^2\delta\left({1\over 4}\right)\right]^{-{1\over 4}p^2} u(p) \\
 V_{ut^{\prime 2}} &=& 2\kappa{\cal R}^{11/2} t^{\prime 2} 
    \int{d^2p\over (2\pi)^2}
    \left[{\cal R}^2\delta\left({1\over 2}\right)\right]^{-{1\over 4}p^2} u(p) \\
 V_{(t^2 + t^{*2})t'} &=& {\sqrt{2}\kappa\over\pi} 
    {\Gamma^2\left({3\over 4}\right)\over\Gamma\left({1\over 2}\right)}
    {\cal R}^{43/8} (t^2 + t^{*2}) t' \;,
\end{eqnarray}
and $\Gamma(1/2)=\sqrt{\pi}$, $\Gamma(3/4)=1.22542$.
Following \cite{OZ} we eliminate $\kappa$ by 
rescaling quantities as follows
\be
p_a&=&2\xi_a,\quad x_a=
\frac{1}{2}r_a, \quad a=1,2\nonumber\\
u(\xi)&=&4\kappa u(p),\quad t\to
\frac{1}{\kappa}t,\quad t'\to\frac{1}{\kappa}t' \;,
\ee
and defining the normalized dimensionless potential 
as \footnote{Note that $f_N$ and $\widetilde {f}_N$ defined in (\ref{f_def}) 
are related as $f_N=N\widetilde {f}_N$ because of (\ref{Newton_rel}).}
\be
f_4(u,t,t')\equiv \frac{\kappa^2 V_4(u,t,t')}{2\pi(1-\frac{1}{4})} \;.
\ee
This gives
\begin{eqnarray}
\label{Potential}
f_4(u,t,t') &=& -\frac{7}{6\pi}t^{\prime 2}-\frac{29}{12\pi}t^*t-\frac{1}{3\pi}
\int\frac{d^2\xi}{(2\pi)^2}u(-\xi)(1-\xi^2)u(\xi)\nonumber\\
&& + {8\over 3\pi} {\cal R}^{45/8} t^* t \int\frac{d^2\xi}{(2\pi)^2}
     \left[{\cal R}^2\delta\left({1\over 2}\right)\right]^{-\xi^2} u(\xi) \nonumber\\
&& + {4\over 3\pi} {\cal R}^{11/2} t^{\prime 2} \int\frac{d^2\xi}{(2\pi)^2}
     \left[{\cal R}^2\delta\left({1\over 4}\right)\right]^{-\xi^2} u(\xi) \nonumber\\
&& +  \beta (t^2 + t^{*2}) t' \nonumber \\
&& + \frac{1}{9\pi}\int\prod_{i=1}^3\left[\frac{d^2\xi_i}{(2\pi)^2}
{\cal R}^{2(1-\xi_i^2)}u(\xi_i)\right](2\pi)^2\delta^{(2)}(\xi_1+\xi_2+\xi_3) \;,
\end{eqnarray}
where
\be
\beta \equiv {2\sqrt{2}\over 3\pi^2} 
     {\Gamma^2\left({3\over 4}\right)\over\Gamma\left({1\over 2}\right)}
     {\cal R}^{43/8} = 0.330244 \;.
\ee
The predicted values at the critical points are 
\be
f_4=4\widetilde {f}_4=-4, -4/3, -4/9 \;,
\label{predicted}
\ee
corresponding to flat space, $\mathbb{C}/\mathbb{Z}_2$ and 
$\mathbb{C}/\mathbb{Z}_3$, respectively.

In our lowest-level approximation we have only included the ground state
tachyons from the twisted sectors. It is therefore reasonable
to include only bulk tachyon modes up to the level of the higher 
of the two twisted tachyons. This imposes a cutoff on the momentum
integrals of the bulk tachyon given by $\xi_* = \sqrt{\ell_{t'}/2} = 1/(2\sqrt{2})$.


\section{The critical points}

Varying the potential (\ref{Potential}) with respect to the bulk and twisted tachyons
we obtain
\begin{eqnarray}
u(\xi)&=&\frac{1}{1-\xi^2}\left\{2{\cal R}^{\frac{11}{2}}t^{\prime 2}\left[{\cal R}^2
\delta\left(\frac{1}{2}\right)\right]^{-\xi^2}+4{\cal R}^{\frac{45}{8}}t^*t
\left[{\cal R}^2\delta\left(\frac{1}{4}\right)\right]^{-\xi^2}\right\}\nonumber\\
&& \mbox{} + \frac{1}{8\pi^2}\frac{{\cal R}^{2(1-\xi^2)}}{1-\xi^2}
\int d^2\xi'{\cal R}^{2(2-\xi'^2-(\xi+\xi')^2)}u(\xi')u(-\xi'-\xi)
\label{u_condition}\\
t' &=& \frac{8}{7}t'{\cal R}^{\frac{11}{2}}
\int\frac{d\xi^2}{(2\pi)^2}\left[{\cal R}^{2}
\delta\left(\frac{1}{2}\right)\right]^{-\xi^2}u(\xi)
  + {3\pi\over 7}\beta (t^2+(t^*)^2) 
\label{t'_condition}\\
t &=& \frac{32}{29}t{\cal R}^{\frac{45}{8}}
\int\frac{d\xi^2}{(2\pi)^2}\left[{\cal R}^{2}\delta\left(\frac{1}{4}\right)\right]^{-\xi^2}
u(\xi) + {24\pi\over 29}\beta  t^*t' 
\label{t_condition}\;.
\end{eqnarray}
Before turning to the solutions let us derive some general properties of  
some of the critical points from these conditions.
The trivial solution $u=t=t'=0$ is the maximum and corresponds
to the original $\mathbb C/\mathbb Z_4$.
If $t'=0$ (\ref{t'_condition}) gives $t^2+(t^*)^2=0$ 
and therefore $t=|t|(\pm 1 \pm i)/\sqrt{2}$.
These four points are related by the $\mathbb Z_4$ symmetry of the potential.
If both $t$ and $t'$ are non-vanishing (\ref{t_condition}) implies
that $t$ must be either real or imaginary.

Solving these conditions analytically is impossible due to the non-linear
integral term in the equation for $u$, so we must resort to some sort
of approximation method. One approach is to discretize the spectrum
of the bulk tachyon by putting it in a box, and then solve the equations
numerically, keeping a finite number of modes. This was the method employed
in \cite{OZ}.
Our approach will be to solve the equations perturbatively, treating
the quadratic term in $u$ as a small perturbation.
In comparing this approach to the one of \cite{OZ} for 
$\mathbb C/\mathbb Z_2$ and $\mathbb C/\mathbb Z_3$, we find similar
results.

To simplify notation we define
\be
v(\xi)&\equiv&{\cal R}^{-2\xi^2}u(\xi)\quad\quad \lambda\equiv 
\frac{{\cal R}^{6}}{8\pi^2}=0.060861 \\
\gamma(\xi)&\equiv&2{\cal R}^{\frac{11}{2}}t^{\prime 2}
\left[\delta\left(\half\right)\right]^{-\xi^2}
 + 4{\cal R}^{\frac{45}{8}}t^*t 
   \left[\delta\left(\frac{1}{4}\right)\right]^{-\xi^2} \;.
\ee
The equation for the bulk tachyon (\ref{u_condition}) then becomes
\be\label{Condition_u}
v(\xi)&=&\frac{{\cal R}^{-4\xi^2}}{1-\xi^2}\left[\gamma(\xi)
+\lambda\int d^2\xi'v(\xi')v(\xi'+\xi)\right] \;.
\ee
Note that only $\gamma(\xi)$ depends on the specific orbifold.
We would like to solve for $v(\xi)$ in terms of the twisted tachyons
$t$ and $t'$ as a perturbative series in $\lambda$,
\be
v(\xi,\lambda)=\sum_{n=0}^{\infty}\frac{\lambda^n}{n!}v^{(n)}(\xi)
\quad\quad v^{(n)}(\xi)=\frac{\partial^n}{\partial\lambda^n}v(\xi,\lambda)\Big|_{\lambda=0}.
\ee
At zeroth order
\be
 v^{(0)}(\xi) = \frac{{\cal R}^{-4\xi^2}}{1-\xi^2}\gamma(\xi)\;,
\ee
and the higher coefficients satisfy the recursion relation
\be
v^{(n)}(\xi) = n\frac{{\cal R}^{-4\xi^2}}{1-\xi^2}\sum_{k=0}^{n-1}\begin{pmatrix}
  n-1\\
  k
\end{pmatrix}
\int d^2\xi'v^{(k)}(\xi')v^{(n-k-1)}(\xi+\xi')\;.
\ee
This can be also seen as an expansion in the number of
integrals. The next two terms in the expansion are given by
\begin{eqnarray}
v^{(1)}(\xi)&=&\frac{{\cal R}^{-4\xi^2}}{1-\xi^2}\int d^2\xi'
\frac{\gamma(\xi')\gamma(\xi'+\xi){\cal R}^{-4(\xi'^2+
(\xi+\xi')^2)}}{(1-\xi'^2)(1-(\xi'+\xi)^2)} \\[5pt]
v^{(2)}(\xi)&=&\frac{2{\cal R}^{-4\xi^2}}{1-\xi^2}\int
\frac{d^2\omega\, d^2\rho\,\gamma(\omega)\gamma(\rho)
\gamma(\omega+\rho+\xi){\cal R}^{-4(\omega^2+\rho^2+
(\rho+\xi)^2+(\omega+\rho+\xi)^2)}}{(1-\omega^2)
(1-\rho^2)(1-(\xi+\rho)^2)(1-(\omega+\rho+\xi)^2)}. 
\end{eqnarray}
Therefore we can expand the effective potential for the twisted tachyons
as 
a power series in $\lambda$ as well,
\be
 f_4(t,t') = \sum_{n=0}^\infty \lambda^n f_4^{(n)}(t,t') \;.
\ee
At zeroth order the cubic bulk tachyon term doesn't contribute and we get 
\be
\label{zero_potential}
f_4^{(0)}(t,t')&=&-\frac{7}{6\pi}t^{\prime 2}-\frac{29}{12\pi}t^*t+
\beta(t^2+(t^*)^2)t'+
\frac{1}{3\pi}\int \frac{d^2\xi}{(2\pi)^2}\gamma(\xi)v_0(\xi)\\
&=&-\frac{7}{6\pi}t^{\prime 2}-\frac{29}{12\pi}t^*t+\beta(t^2+(t^*)^2)t'+
(at^*t+bt^{\prime 2})^2-ct^*tt^{\prime 2},\nonumber
\ee
where $\beta=0.330244$, $a=0.44631$, $b=0.23264$ and $c=0.00025$.
There are four classes of critical points:
\begin{enumerate}
\item $(t,t')^{(0)}_{\mathbb Z_1}=(\pm 1.801,-1.722), (\pm 1.801i,1.722)$ 
\hfill $f_4^{(0)} = -2.719$
\item $(t,t')^{(0)}_{\mathbb Z_2}=(\pm 0.983 \pm 0.983i,0)$ 
\hfill  $f_4^{(0)} = -0.743$
\item $(t,t')^{(0)}_{\mathbb Z_2'}= (0, \pm1.852)$ \hfill  $f_4^{(0)} = -0.637$
\item $(t,t')^{(0)}_{\mathbb Z_3}=(\pm 0.688,0.718), (\pm 0.688i,-0.718)$ 
\hfill  $f_4^{(0)} = -0.221$
\end{enumerate}
The relative values of the tachyon potential at the critical points suggest
that the first class of points corresponds to flat space,
the second and third to $\mathbb C/\mathbb Z_2$
and the fourth to $\mathbb C/\mathbb Z_3$ (hence the labels).
This can be verified by examining the spectrum of fluctuations around
each kind of critical point, and counting the number of negative modes,
{\em i.e.} tachyons (see appendix B). 
Comparing with the predicted values (\ref{predicted}) 
we find that the minimum is at $68\%$ of its predicted value of $-4$,
the $\mathbb C/\mathbb Z_2$ points are at $55\%$ and $48\%$ 
(for the two classes) of their predicted value of $-4/3$, and
the $\mathbb C/\mathbb Z_3$ points are at $50\%$ of their 
predicted value of $-4/9$.
The results for the first four orders in perturbation
theory are summarized in table \ref{z4pot_vals} (for more details see
appendix A). The depth of the minimum dramatically decreases: we get $25\%$ of the 
predicted value for flat 
space at third order.
The depths of the $\mathbb C/\mathbb Z_2$ and $\mathbb C/\mathbb Z_3$ 
points also decrease with the order of the calculation, but not nearly
as much. We end up with $39\%$ and $34\%$ agreement for the two
classes of $\mathbb C/\mathbb Z_2$ points, and $46\%$ for the
$\mathbb C/\mathbb Z_3$ points.
\begin{table*}[htbp]
\begin{center}
\begin{tabular}{|c|c|c|c|c|}
  \hline
   Order& Vacuum & $\mathbb{C}/\mathbb{Z}_2$ quartet & $\mathbb{C}/\mathbb{Z}_2$ 
doublet & $\mathbb{C}/\mathbb{Z}_3$ \\
   \hline
  0 & -2.71904& -0.742683& -0.637051& -0.221467\\
  \hline
  1 &-1.45297 & -0.587339&  -0.505818& -0.208521\\
  \hline
  2 & -1.18091& -0.543738& -0.46922&-0.205589 \\
  \hline
  3 & -1.06427& -0.52420& -0.452928&-0.204618 \\
   \hline\hline
    Predicted&-4&-4/3&-4/3&-4/9\\
   \hline
\end{tabular}
\caption{The calculated values of $f_4$ in first orders of perturbation series.}
\label{z4pot_vals}
\end{center}
\end{table*}

The degeneracy of the potential within each class of critical points
is a consequence of the $\mathbb Z_4$ symmetry of the orbifold theory.
The critical points are located roughly at the vertices, edges and faces
of a tetrahedron in the three-dimensional twisted-tachyon space (Fig.~1). 
The four vertices correspond to flat space, 
the six edges to $\mathbb C/\mathbb Z_2$, and
the four faces to $\mathbb C/\mathbb Z_3$.
The center of the tetrahedron corresponds to the original orbifold 
$\mathbb C/\mathbb Z_4$. In Fig.~2 we show contour plots of the tachyon
potential on two different two-dimensional sections, revealing the 
tetrahedral structure.
The tetrahedron defined by the four vertices is
not a regular one; it has four edges (two pairs of opposing edges)
of one length, and two of another length.
Its symmetry group is $D_4$, which is a subgroup of the
full symmetry of the regular tetrahedron $T_d$.
The quantum symmetry of the orbifold $\mathbb Z_4$ is the maximal abelian
subgroup of $D_4$.
The four equal edges do not mix with the other two under
$\mathbb Z_4$ or $D_4$, which explains the two classes of 
$\mathbb C/\mathbb Z_2$ points.
Ultimately we know that since there is a unique $\mathbb C/\mathbb Z_2$
orbifold all the $\mathbb C/\mathbb Z_2$ points must be the same.
An interesting question is whether the symmetry of the tachyon potential
gets enhanced to the full tetrahedral group $T_d$ at higher level,
thereby guaranteeing a six-fold degeneracy of $\mathbb C/\mathbb Z_2$ points.

\begin{figure}[htbp]
\centerline{\epsfxsize=2.5in\epsfbox{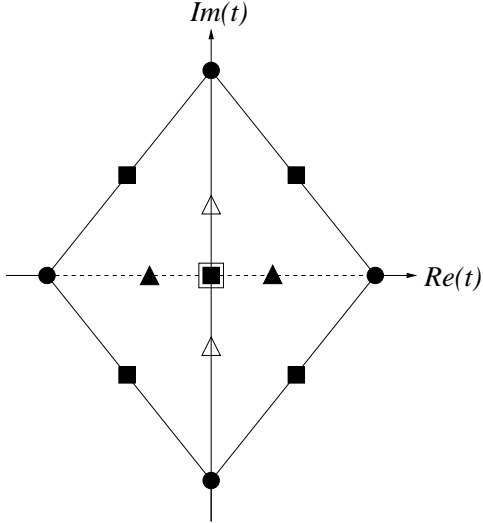}}
\medskip
\caption{A schematic picture of the critical points of $f_4(t,t')$ projected onto the $t$-plane.
The circles are the minima, {\em i.e.} flat space, the squares are the $\mathbb C/\mathbb Z_2$
points, and the triangles are the $\mathbb C/\mathbb Z_3$ points.}
\end{figure}

To check that our approximation scheme is as good as that of
\cite{OZ} we applied it to $\mathbb{C}/\mathbb{Z}_{2}$ and
$\mathbb{C}/\mathbb{Z}_{3}$ as well.
Using the perturbative technique to third order
we find the minimum for $\mathbb{C}/\mathbb{Z}_{2}$ 
with $f_2=-0.6794$, compared to $f_2=-0.7202$ in \cite{OZ}.
For $\mathbb{C}/\mathbb{Z}_{3}$ the perturbative
expansion to third order gives $f_3=-0.9882,-0.3240$ for the
minimum and $\mathbb{C}/\mathbb{Z}_{2}$ points respectively,  
compared to $f_3=-0.9889,-0.3356$ in \cite{OZ}.


\begin{figure}
\epsfxsize=3in\epsfbox{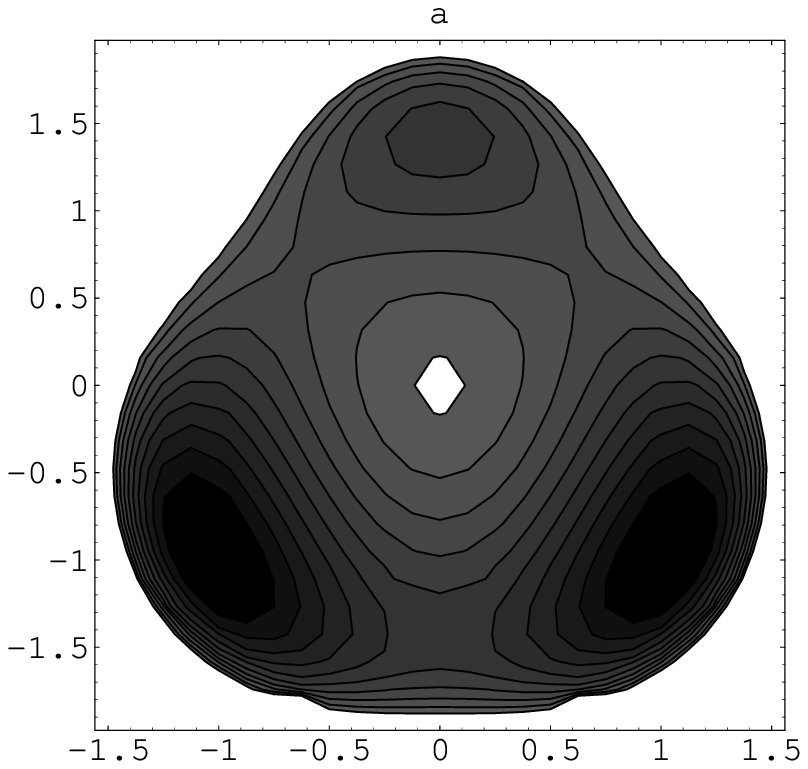}
\epsfxsize=3in\epsfbox{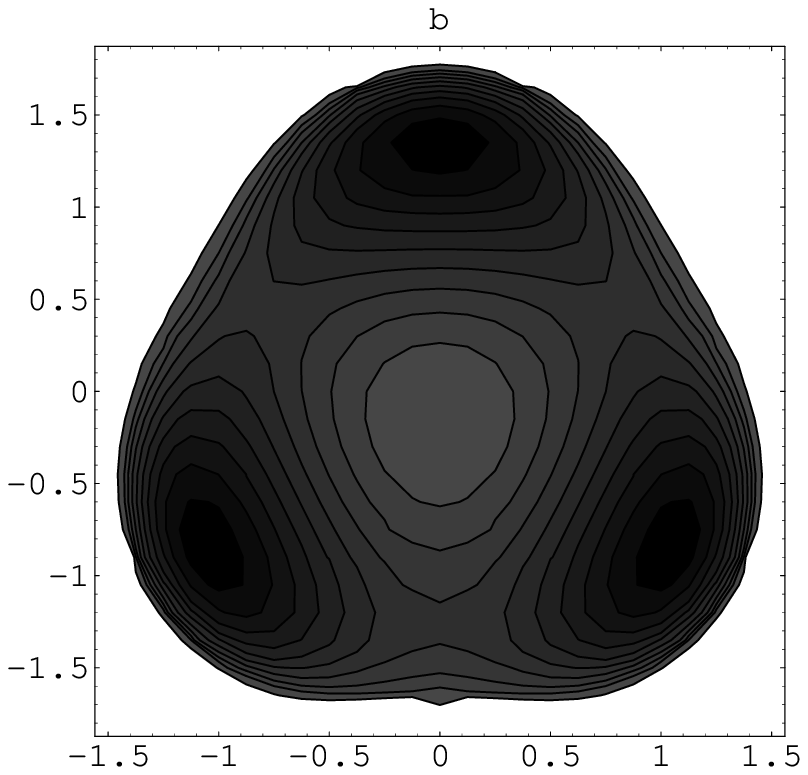}
\medskip
\caption{Contour plots of $f_4(t,t')$ on
(a) the slice $\mbox{Im}(t)=0$, (b) one of the faces of the
tetrahedron.
The first plot shows the maximum $f_4=0$ at the origin,
two of the minima at $(\pm 1.045,-0.970)$, two $\mathbb C/\mathbb Z_2$
saddle points at $(0,\pm 1.441)$ (one is a local minimum
on this cross section), and two $\mathbb C/\mathbb Z_3$ points
at $(\pm 0.639,0.671)$.
The second plot shows (roughly) a $\mathbb C/\mathbb Z_3$ point 
(a local maximum on this cross section) at the origin,
three minima, and three $\mathbb C/\mathbb Z_2$ saddle points.}
\end{figure}


\subsection{A consistency check}

Not all of the critical points are independent.
The $\mathbb C/\mathbb Z_2$ doublet of the $\mathbb C/\mathbb Z_4$
potential is in fact related to the minimum of $\mathbb C/\mathbb Z_2$ 
potential (at lowest level).  
Consider the $\mathbb C/\mathbb Z_4$ tachyon potential with 
the complex tachyon vanishing, $V_4(0,t')$. The real tachyon $t'$ corresponds
to the twist 2 field $\sigma_2$ (the middle twisted sector),
which has the same conformal dimension as the twist 1 field $\sigma_1$
of the $\mathbb C/\mathbb Z_2$ orbifold, since in both cases
$k/N = 1/2$. This means all correlation functions involving this field
and the untwisted fields are equal in the two theories. 
In particular the tachyon potentials are equal \footnote{For the purpose
of comparing the potentials in the two theories we use $t'$ to denote
both the twist 2 field in $\mathbb C/\mathbb Z_4$ and the twist 1
field in $\mathbb C/\mathbb Z_2$.}
\be
 V_4(0,t') = V_2(t') \;.
\label{equal_potentials}
\ee
The minimum of the potential on the right hand side corresponds
to flat space in the  $\mathbb C/\mathbb Z_2$ theory. 
On the other hand minimizing the potential 
on the left hand side with respect to $t'$ gives the $\mathbb C/\mathbb Z_2$
doublet in the $\mathbb C/\mathbb Z_4$ theory.
The equality above could therefore provide an alternative prediction for these
$\mathbb C/\mathbb Z_2$ points. In terms of the normalized potential $\widetilde {f}_4$
the original prediction (\ref{prediction_2}) for (all) the $\mathbb C/\mathbb Z_2$
points is $\widetilde {f}_4=-1/3$, and the one that follows from 
(\ref{equal_potentials}) is $\widetilde{f}'_4=(1/3)\widetilde{f}_2=-1/3$,
so the predictions agree.

We can generalize this observation for higher orbifolds.
Any $\mathbb C/\mathbb Z_N$ orbifold for which $N$ is a product exhibits
potential multiple predictions for some of the critical points.
Consider for example $\mathbb C/\mathbb Z_{pq}$. 
The twist $qk$ field $\sigma_{qk}$ has the same dimensions as
the twist $k$ field in $\mathbb C/\mathbb Z_{p}$.
Therefore setting all tachyons but the ones associated with
the twist fields $\sigma_{qk}$ for $k=1,\ldots,p-1$ to zero,
the tachyon potential $V_{pq}$ becomes the same as the tachyon
potential $V_p$ of the $\mathbb C/\mathbb Z_{p}$ theory,
\be
 V_{pq}(0,\ldots,0,t_q=t'_1,0,\ldots,0,t_{2q}=t'_2,\ldots) = 
V_p(t'_1,t'_2,\ldots) \;.
\ee
We now want to show that the critical points 
with respect to $t_{qk}$ of the reduced potential
on the left hand side are also critical points of
the full potential $V_{pq}$ with respect to all the $t_k$.
The only possible problems can come from terms in the potential
linear in one of the vanishing tachyons,
$t_s\prod_{i=1}^{n-1}t_{qk_i}$, where $s$ is not an integer
multiple of $q$.
Varying with respect to $t_s$ 
one could arrive at a non vanishing result for 
the vanishing tachyon $t_s$. However such terms are 
forbidden by the quantum symmetry of the orbifold $Z_{pq}$,
since this requires $s+ q\sum_i k_i=npq$ for some 
integer $n$, and therefore $s= q(np-\sum_i k_i)$.
Thus the critical points of $V_{pq}$ with $t_s=0$ ($s\neq qk$)
are identical to the critical points of $V_p$.
This provides a possible alternative prediction for each of these critical
points of $V_{pq}$. The points in question correspond to the orbifolds
$\mathbb C/\mathbb Z_{kq}$ with $k=1,\ldots p-1$.
The original prediction for the normalized potential (\ref{prediction_2}) is
\be
 \widetilde {f}_{pq}(\mathbb C/\mathbb Z_{kq}) = 
-{{1\over k}-{1\over p}\over q - {1\over p}}\;,
\ee
and the one that follows from the relation to $V_p$ is
\be 
 \widetilde{f}'_{pq}(\mathbb C/\mathbb Z_{kq}) = 
 {\kappa^2_{pq}\over \kappa^2_p} {1-{1\over p}\over 1 - {1\over pq}}
  \widetilde {f}_p(\mathbb C/\mathbb Z_{k}) = 
 -{{1\over k}-{1\over p}\over q - {1\over p}}   \;,
\ee
in agreement with $\widetilde{f}_{pq}(\mathbb C/\mathbb Z_{kq})$.


\section{Conclusions}

Extending the work of \cite{OZ}, we have computed the potential
for localized tachyons in the bosonic string orbifold $\mathbb{C}/\mathbb{Z}_4$ 
using a lowest-level truncation of closed string field theory to cubic order.
Our computation provides additional evidence for the conjecture of 
\cite{APS}, that the critical points of the tachyon potential 
correspond to lower-order orbifolds, and in particular that the minimum
is flat space. 
The results are consistent with those of \cite{OZ} for the 
$\mathbb{C}/\mathbb{Z}_2$ and $\mathbb{C}/\mathbb{Z}_3$ theories.
We also believe that these results can be taken as support
for the general method of level-truncation, together with string-vertex
truncation, in closed string field theory.

There are a couple of interesting puzzles that require further study.
The first is that there appear to be two classes of $\mathbb{C}/\mathbb{Z}_2$ 
critical points, with different values of the tachyon potential.
The $\mathbb Z_4$ (or $D_4$) symmetry of the tachyon potential guarantees the
degeneracy within each class, but not between the two classes.
The latter requires the full tetrahedral symmetry $T_d$.
The question is whether this symmetry is restored at higher level,
or whether at higher level the two classes of points become
accidentally degenerate.
The former scenario is more appealing, since it would 
also explain the quantum symmetries of the different critical points.
The tetrahedral group $T_d$ contains $\mathbb Z_4, \mathbb Z_3$ and $\mathbb Z_2$,
corresponding to the abelian subgroups preserving the center, a face, and a pair of
opposite edges, respectively. We could therefore identify these groups
as the unbroken symmetries of $\mathbb{C}/\mathbb{Z}_4$, $\mathbb{C}/\mathbb{Z}_3$ 
and $\mathbb{C}/\mathbb{Z}_2$, respectively.

Another interesting puzzle raised by the calculation of localized closed string 
tachyon potentials is the possibility of stable non-uniform tachyon configurations,
{\em i.e.} tachyonic solitons.
For example in the $\mathbb{C}/\mathbb{Z}_2$ theory the effective potential for
the twisted tachyon has two degenerate minima \cite{OZ}, and should therefore admit
a kink solution.
In the analogous situation for open strings, namely an unstable (non-BPS) D$p$-brane
in Type II (or Type 0) string theory,
the tachyonic kink corresponds to a stable (BPS) D$(p-1)$-brane \cite{Sen_solitons,Sen_review}.
The topological stability of the kink configuration in the orbifold implies that it corresponds
to a closed string background which is stable (modulo the usual bulk tachyon).
This appears to be a new background of the bosonic string, which is as stable
as flat space. It would be interesting to study this background further.

\section*{Acknowledgments}
We would like to thank Alex Flournoy, Shinji Hirano, 
Yuji Okawa and Barton Zwiebach for useful discussions.
This work is supported in part by the
Israel Science Foundation under grant no.~101/01-1.

\appendix
\section{Details of the perturbative calculation}

Here we give the details of the calculation of the first few orders in
the perturbative expansion of the twisted tachyon effective potential. 
Recall that the first couple of terms in the expansion for the
scaled bulk tachyon $v(\xi)$ are given by
\be
v^{(1)}(\xi)&=&\frac{v^{(0)}(\xi)}{\gamma(\xi)}\int d^2\xi_1v^{(0)}(\xi_1)v^{(0)}(\xi_1+\xi),\\
v^{(2)}(\xi)&=&4\frac{v^{(0)}(\xi)}{\gamma(\xi)}\int d^2\xi_1v^{(1)}(\xi_1)v^{(0)}(\xi_1+\xi)\nonumber
\ee
The tachyon potential is expanded as
\be
f_4=\sum_{k=0}^\infty\lambda^kf_4^{(k)}.
\ee
The zeroth order term was given in the bulk of the paper (\ref{zero_potential}).
The first order term is given by
\be
f^{(1)}_4(t,t')&=&\frac{8\pi}{9}\int\frac{d^2\xi_1d^2\xi_2}{(2\pi)^4}
v^{(0)}(\xi_1)v^{(0)}(\xi_2)v^{(0)}(\xi_1+\xi_2)\\
&=&a^{(1)}_{30}(tt^*)^3+a^{(1)}_{03}t'^6+a^{(1)}_{21}(tt^*)^2t'^2+a^{(1)}_{12}t'^4(tt^*),\nonumber
\ee
where
\be
a^{(1)}_{12}=\frac{6(2{\cal R}^{11/2})^2(4{\cal R}^{45/8})}{9\pi}\eta(1/2,1/4)
=0.4846\;,& a^{(1)}_{30}=\frac{2(4{\cal R}^{45/8})^3}{9\pi}\eta(1/4,1/4)=0.5940\nonumber\\
a^{(1)}_{21}=\frac{6(4{\cal R}^{45/8})^2(2{\cal R}^{11/2})}{9\pi}\eta(1/4,1/2)=0.9750\;,&
a^{(1)}_{03}=\frac{2(2{\cal R}^{11/2})^3}{9\pi}\eta(1/2,1/2)=0.0887\nonumber
\ee
and where we have defined
\be
\eta(a,b)\equiv\int\frac{d^2\xi d^2\xi'}{(2\pi)^2}\frac{({\cal R}^4
\delta(a))^{-\xi^2-\xi'^2}({\cal R}^4\delta(b))^{-(\xi'+\xi)^2}}{(1-\xi^2)(1-\xi'^2)(1-(\xi'+\xi)^2)}\;.
\ee
The second order term is given by
\be
f^{(2)}_4(t,t')&=&-\frac{1}{3\pi}\int\frac{d^2\xi}{(2\pi)^2}
\frac{\gamma(\xi)}{v^{(0)}(\xi)}\left(v^{(1)}(\xi)\right)^2+\frac{8\pi}{3}\int
\frac{d^2\xi_1d^2\xi_2}{(2\pi)^4}v^{(1)}(\xi_1)v^{(0)}(\xi_2)v^{(0)}(\xi_1+\xi_2)\nonumber\\
&=&\frac{1}{3\pi}\int\frac{d^2\xi}{(2\pi)^2}\frac{\gamma(\xi)}{v^{(0)}(\xi)}\left(v^{(1)}(\xi)\right)^2\\
&=&a^{(2)}_{40}(tt^*)^4+a^{(2)}_{04}t'^8+a^{(2)}_{31}(tt^*)^3t'^2
+a^{(2)}_{13}t'^6(tt^*)+a^{(2)}_{22}t'^4(tt^*)^2,\nonumber
\ee
where
\be
a^{(2)}_{13}&=&\frac{(2{\cal R}^{11/2})^3(4{\cal R}^{45/8})}{3\pi}\tilde\eta_1(1,3)=2.4433\;,
\; a^{(2)}_{40}=\frac{(4{\cal R}^{45/8})^4}{3\pi}\tilde\eta_1(4,0)=3.9897\nonumber\\
a^{(2)}_{31}&=&\frac{(4{\cal R}^{45/8})^3(2{\cal R}^{11/2})}{3\pi}\tilde\eta_1(3,1)=8.5329\;,\;
a^{(2)}_{04}=\frac{(2{\cal R}^{11/2})^4}{3\pi}\tilde\eta_1(0,4)=0.3271\nonumber \\
a^{(2)}_{22}&=&\frac{(2{\cal R}^{11/2})^2(4{\cal R}^{45/8})^2}{3\pi}\tilde\eta_1(2,2)=6.8472\;,\nonumber
\ee
and where we have defined
\be
\tilde \eta_1(k,l)&\equiv& \frac{1}{k!l!}\sum_{P\in S_4}\eta\Big(P(\underbrace{1/4,\ldots,1/4}_{k},
\underbrace{1/2,\ldots,1/2}_{l})\Big)\;,\quad k+l=4\;,\\
\eta(a,b,c,d)&\equiv& \int\frac{d^2\xi\, d^2\omega\, d^2\rho}{(2\pi)^2}
\frac{({\cal R}^4\delta(a))^{-\omega^2}({\cal R}^4\delta(c))^{-(\omega+\xi)^2}
({\cal R}^4\delta(b))^{-\rho^2}({\cal R}^4\delta(d))^{-(\rho+\xi)^2}
{\cal R}^{-4\xi^2}}{(1-\xi^2)(1-\omega^2)(1-(\omega+\xi)^2)(1-\rho^2)
(1-(\rho+\xi)^2)}.\nonumber
\ee
The third order term is given by
\be
f^{(3)}_4(t,t')&=&-\frac{1}{3\pi}\int\frac{d^2\xi}{(2\pi)^2}
\frac{\gamma(\xi)}{v^{(0)}(\xi)}v^{(1)}(\xi)v^{(2)}(\xi)
+ \frac{8\pi}{3}\int\frac{d^2\xi_1d^2\xi_2}{(2\pi)^4}v^{(1)}(\xi_1)v^{(1)}(\xi_2)v^{(0)}(\xi_1+\xi_2)\nonumber\\
&+&\frac{4\pi}{3}\int\frac{d^2\xi_1d^2\xi_2}{(2\pi)^4}v^{(2)}(\xi_1)v^{(0)}(\xi_2)v^{(0)}(\xi_1+\xi_2)\\
&=&\frac{1}{6\pi}\int\frac{d^2\xi}{(2\pi)^2}v^{(1)}(\xi)v^{(2)}(\xi)
\frac{\gamma(\xi)}{v^{(0)}(\xi)}\nonumber\\
&=&a^{(3)}_{50}(tt^*)^5+a^{(3)}_{05}t'^{10}+a^{(3)}_{41}(tt^*)^4t'^2+a^{(3)}_{14}t'^8(tt^*)
+a^{(3)}_{32}t'^4(tt^*)^3+a^{(3)}_{23}t'^6(tt^*)^2,\nonumber
\ee
where
\be
a^{(3)}_{14}&=&\frac{2(2{\cal R}^{11/2})^4(4{\cal R}^{45/8})}{3\pi}\tilde\eta_2(1,4)=14.9425\;,\;
a^{(3)}_{50}=\frac{2(4{\cal R}^{45/8})^5}{3\pi}\tilde\eta_2(5,0)=35.7330\nonumber\\
a^{(3)}_{41}&=&\frac{2(4{\cal R}^{45/8})^4(2{\cal R}^{11/2})}{3\pi}\tilde\eta_2(4,1)=96.0050\;,\;
a^{(3)}_{05}=\frac{2(2{\cal R}^{11/2})^5}{3\pi}\tilde\eta_2(0,5)=1.6079\nonumber \\
a^{(3)}_{23}&=&\frac{2(2{\cal R}^{11/2})^3(4{\cal R}^{45/8})^2}{3\pi}\tilde\eta_2(2,3)=55.5283\nonumber\\
a^{(3)}_{32}&=&\frac{2(2{\cal R}^{11/2})^2(4{\cal R}^{45/8})^3}{3\pi}\tilde\eta_2(3,2)=103.2366\nonumber
\ee
and where we have defined
\be
\tilde \eta_2(k,l)&\equiv& \frac{1}{k!l!}\sum_{P\in S_5}\eta\Big(P(\underbrace{1/4...1/4}_{k},
\underbrace{1/2...1/2}_{l})\Big),\quad k+l=5\\
\eta(a,b,c,d,e)&=&\int\frac{d^2\xi\, d^2\xi'\, d^2\omega\, d^2\rho}{(2\pi)^2}\nonumber\\
&\times&\frac{({\cal R}^4\delta(a))^{-\rho^2}({\cal R}^4\delta(b))^{-(\xi+\omega)^2}
({\cal R}^4\delta(c))^{-\omega^2}({\cal R}^4\delta(d))^{-(\xi+\xi')^2}
({\cal R}^4\delta(e))^{-(\xi'+\rho)^2}{\cal R}^{-4(\xi^2+\xi'^2)}}{(1-\xi^2)(1-\xi'^2)
(1-(\xi+\xi')^2)(1-\omega^2)(1-\rho^2)(1-(\xi+\omega)^2)(1-(\xi'+\rho)^2)}.\nonumber
\ee
The positions and potential values of the critical points at each order
are shown in table \ref{pert_points}.
\begin{table*}[htbt]
\begin{tabular}{|c|c|c|c|}
 \hline
 \multicolumn{4}{|c|}{flat space} \\ \hline
  order & $t$ & $t'$ & $f_4$ \\\hline
  0 & 1.8006 & -1.7216 & -2.7190\\\hline
  1 & 1.2708 & -1.1994 & -1.4530\\\hline
  2 & 1.1192 & -1.0452 & -1.1809\\\hline
  3 & 1.0451 & -0.9696 & -1.0643\\\hline
\end{tabular}
\qquad
\begin{tabular}{|c|c|c|c|}
 \hline
 \multicolumn{4}{|c|}{$\mathbb{C}/\mathbb{Z}_3$} \\ \hline
  order & $t$ & $t'$ & $f_4$\\\hline
  0 & 0.6876 & 0.7177 & -0.2215 \\\hline
  1 & 0.6533 & 0.6844 & -0.2085\\\hline
  2 & 0.6434 & 0.6751 & -0.2056\\\hline
  3 & 0.6394 & 0.6713 & -0.2046\\\hline
\end{tabular}\\[20pt]
\begin{tabular}{|c|c|c|c|}
 \hline
 \multicolumn{4}{|c|}{$\mathbb{C}/\mathbb{Z}_2$ doublet} \\ \hline
  order & $t$ & $t'$ & $f_4$ \\\hline
  0 & 0 & 1.8523 & -0.6371\\\hline
  1 & 0 & 1.5804 & -0.5058\\\hline
  2 & 0 & 1.4876 & -0.4692\\\hline
  3 & 0 & 1.4407 & -0.4529\\\hline
\end{tabular}
\qquad\qquad\;
\begin{tabular}{|c|c|c|c|}
  \hline
  \multicolumn{4}{|c|}{$\mathbb{C}/\mathbb{Z}_2$ quartet} \\ \hline
  order & $t$ & $t'$ &$f_4$ \\\hline
  0 & 0.9826(1+$i$) & 0 & -0.7427\\\hline
  1 & 0.8362(1+$i$) & 0 & -0.5873\\\hline
  2 & 0.7860(1+$i$) & 0 & -0.5437\\\hline
  3 & 0.7604(1+$i$) & 0 & -0.5242\\\hline
\end{tabular}
\caption{Extremal points of the tachyon potential $f_4$ at each order of
perturbation theory. There are additional points related to
these by the $\mathbb{Z}_4$ symmetry.}
\label{pert_points}
\end{table*}


\section{Fluctuation analysis at the critical points}

Here we will analyze the spectrum of fluctuations around the critical points
of the tachyon potential. In particular we will show that the number of tachyons
agrees with the identification of each critical point with a particular orbifold.
For simplicity we look only at the zeroth order potential (\ref{zero_potential})
(this should not affect the number of tachyons, only the precise values of their masses).
The quadratic terms in the fluctuations of the tachyons $\delta t$, $\delta t'$ and $\delta u$
are given by
\be
f^{(0)}_4(t+\delta t,t'+\delta t',u(t,t')+\delta u)&=&
\left[-\frac{7}{6\pi}+(2ab-c)|t|^2+6b^2t'^2\right]
\delta t'^2 \\
&+&  \quart\left[-\frac{29}{12\pi}+(2ab-c)t'^2 + \half a^2(3t_+^2+t_-^2) + 2\beta t'\right]\delta t_+^2\nonumber\\
&+& \quart\left[-\frac{29}{12\pi}+(2ab-c)t'^2+\half a^2(3t_-^2+t_+^2) - 2\beta t'\right]\delta t_-^2
\nonumber\\
&+&  \Big[(2ab-c)t_+ t' + \beta t_+\Big] \delta t_+ \delta t' \nonumber \\
&+& \Big[(2ab-c)t_- t' - \beta t_-\Big] \delta t_- \delta t' \nonumber\\
&+& \half a^2t_+t_-\delta t_+\delta t_- 
- \frac{1}{3\pi}\int\frac{d^2\xi}{(2\pi)^2}\delta u(-\xi)(1-\xi^2)\delta u(\xi)\;,\nonumber
\ee
where $t_+\equiv 2\mbox{Re}(t)$ and $t_-\equiv 2\mbox{Im}(t)$.
We see that there is always at least one tachyonic mode coming from the bulk
tachyon $\delta u$.
Let us analyze each of the four classes of critical points in turn.
\begin{enumerate}
\item \underline{Minimum}: $(t,t')=(1.801,-1.722)$.
Diagonalizing the mass matrix of the fluctuations we find
\be
 m_1^2=0.4071\;,\;
 m_2^2=0.1572\;,\;
 m_3^2=1.2678 \;,
\ee
corresponding respectively to $\delta t_-$, $(0.998\delta t_+ + 0.062\delta t')$
and $(-0.062\delta t_+ + 0.998\delta t')$.
There are no tachyons (other than the bulk tachyon), which is consistent with the identification of this point
with flat space.
\item \underline{$\mathbb C/\mathbb Z_2$ quartet}: $(t,t')=(0.983(1+i),0)$.
The eigenvalues of the mass matrix are given by
\be
 m_1^2 = -1.0060\;,\;
 m_2^2 = 0.8429\;,\;
 m_3^2 = 0.5769
\ee
corresponding to the eigenvectors $(-0.529 \delta t_++0.529\delta t_-+0.663\delta t')$,
$(0.469 \delta t_+-0.469\delta t_-+0.748\delta t')$ and $(0.707 \delta t_++0.707\delta t_-)$.
There is a single tachyonic mode, as one expects from the twisted sector of $\mathbb C/\mathbb Z_2$.
\item \underline{$\mathbb C/\mathbb Z_2$ doublet}: $(t,t')=(0,1.852)$.
Here there is no mixing between $\delta t$ and $\delta t'$, and we simply get
\be 
m^2_{\delta t'}=0.7427\;,\; 
m^2_{\delta t_+}=0.2914\;,\; 
m^2_{\delta t_-}=-0.3203 \;,
\ee
Again there is just one tachyon, which is consistent with $\mathbb C/\mathbb Z_2$.
\item \underline{$\mathbb C/\mathbb Z_3$}: $(t,t')=(0.688,0.718)$.
Here we get 
\be
 m_1^2=-0.2606\;,\;
 m_2^2=0.4291\;,\;
 m_3^2=-0.5116 \;,
\ee
corresponding respectively to $\delta t_-$, $(0.754 \delta t_++0.657\delta t')$ and 
$(-0.657 \delta t_++0.754\delta t')$.
We find two tachyonic modes, as there should be for $\mathbb C/\mathbb Z_3$.
\end{enumerate}

\end{document}